\documentclass{PoS}

\usepackage{slashed}

\title{Light-front quantization is the same as instant-time quantization}
\ShortTitle{Light-front quantization}
\author{\speaker{Philip D. Mannheim} \\
Department of Physics, University of Connecticut, Storrs, CT 06269, USA\\
\email{philip.mannheim@uconn.edu}}
\date{December 22, 2019}

\abstract{
Commutation or anticommutation relations quantized at equal instant time and commutation or anticommutation relations quantized at equal light-front time cannot be transformed into each other. While they would thus appear to describe different theories, we show that this is not in fact the case. In instant-time quantization unequal instant-time commutation or anticommutation relations for free scalar, fermion, or gauge boson fields are c-numbers. We show that when these unequal instant-time commutation or anticommutation relations are evaluated at equal light-front time they are identical to the equal light-front time commutation or anticommutation relations. Light-front quantization and instant-time quantization are thus the same and thus describe the same physics.}

\FullConference{Light Cone 2019 - QCD on the light cone: from hadrons to heavy ions\\
		16-20 September 2019\\
		Paris, France}

\begin{document}

\section{Introduction}
\label{S1}

In quantum field theory various choices of quantization are considered. The most common choice is to take commutation relations of pairs of fields at equal instant time $x^0$ to be specific singular c-number functions. Thus for a free massless scalar field with action 
\begin{eqnarray}
I_S=\int dx^0dx^1dx^2dx^3\textstyle{\textstyle{\frac{1}{2}}}\partial_{\mu}\phi\partial^{\mu}\phi
=\int dx^0dx^1dx^2dx^3\textstyle{\frac{1}{2}}[(\partial_0\phi)^2-(\partial_1\phi)^2-(\partial_2\phi)^2-(\partial_3\phi)^2]
\label{Q1}
\end{eqnarray}
for instance,  one identifies a canonical conjugate $\delta I_S/\delta \partial_0\phi=\partial^0\phi=\partial_0\phi$  and then quantizes the theory according to the equal-time canonical commutation relation
\begin{eqnarray}
[\phi(x^0,x^1,x^2,x^3), \partial_0\phi(x^0,y^1,y^2,y^3)]=i\delta(x^1-y^1)\delta(x^2-y^2)\delta(x^3-y^3).
\label{Q2}
\end{eqnarray}
In light-front quantization (see e.g. \cite{Brodsky:1997de} for a review) one introduces coordinates $x^{\pm}=x^0\pm x^3$ and metric $x^+x^--(x^1)^2-(x^2)^2=g_{\mu\nu}dx^{\nu}dx^{\nu}$ with $(-g)^{1/2}=1/2$. With the action now of the form
\begin{eqnarray}
I_S&=&\textstyle{\frac{1}{2}}\int dx^+dx^1dx^2dx^-\textstyle{\frac{1}{2}}[2\partial_+\phi\partial_-\phi+2\partial_-\phi\partial_+\phi-(\partial_1\phi)^2-(\partial_2\phi)^2],
\label{Q4}
\end{eqnarray}
one identifies a canonical conjugate $(-g)^{-1/2}\delta I_S/\delta \partial_+\phi=\partial^+\phi=2\partial_-\phi$, and quantizes the theory according to the equal light-front time $x^+$ commutation relation (see e.g. \cite{Mannheim2019a} and references therein)
\begin{eqnarray}
[\phi(x^+,x^1,x^2,x^-), 2\partial_-\phi(x^+,y^1,y^2,y^-)]=i\delta(x^1-y^1)\delta(x^2-y^2)\delta(x^--y^-).
\label{Q5}
\end{eqnarray}
As written, (\ref{Q5}) is already conceptually different from (\ref{Q2}) since the light-front conjugate is $2\partial_-\phi$ and not $2\partial_+\phi$, i.e., not the derivative with respect to the light-front time, while the instant-time conjugate $\partial_0\phi$ is the derivative with respect to the instant time. Since $\phi(x^+,x^1,x^2,x^-)$ and $\partial_-\phi(x^+,y^1,y^2,y^-)$ are not at the same $x^-$, (\ref{Q5}) can be integrated to
\begin{eqnarray}
[\phi(x^+,x^1,x^2,x^-), \phi(x^+,y^1,y^2,y^-)]=-\textstyle{\frac{i}{4}}\epsilon(x^--y^-)\delta(x^1-y^1)\delta(x^2-y^2),
\label{Q6}
\end{eqnarray}
where $\epsilon(x)=\theta(x)-\theta(-x)$. Since the analog instant-time commutation relation  is given by
\begin{eqnarray}
[\phi(x^0,x^1,x^2,x^3), \phi(x^0,y^1,y^2,y^3)]=0,
\label{Q7}
\end{eqnarray}
instant-time and light-front quantization appear to be quite different. Nonetheless, as shown in \cite{Mannheim2019a,Mannheim2019b} instant-time and light-front time matrix elements such as $\langle \Omega |T[\phi(x)\phi(y)]|\Omega\rangle$ (as time ordered with $x^0$ or $x^+$) are actually equal, and in this sense the two quantization schemes are equivalent.

In the present paper we establish an equivalence between the two quantization schemes at the operator level itself without needing to take matrix elements. To this end we note that in instant-time quantization one can use the equal-time commutation relation given in (\ref{Q2}) and the wave equation $\partial_{\mu}\partial^{\mu}\phi=0$ associated with $I_S$ to make an on-shell Fock space expansion of $\phi$ of the form
\begin{eqnarray}
\phi(\vec{x},x^0)=\int \frac{d^3p}{(2\pi)^{3/2}(2p)^{1/2}}[a(\vec{p})e^{-ip x^0+i\vec{p}\cdot\vec{x}}+a^{\dagger}(\vec{p})e^{+ip x^0-i\vec{p}\cdot\vec{x}}],
\label{Q8}
\end{eqnarray}
where the normalization of the creation and annihilation operator algebra, viz.  $[a(\vec{p}),a^{\dagger}(\vec{q})]=\delta^3(\vec{p}-\vec{q})$
is fixed from the normalization of the canonical commutator given in (\ref{Q2}). Given (\ref{Q8}) one can evaluate the unequal-time commutation relation between two free scalar fields, to obtain
\begin{eqnarray}
&&i\Delta(x-y)=[\phi(x^0,x^1,x^2,x^3), \phi(y^0,y^1,y^2,y^3)]
\nonumber\\
&&=
\int \frac{d^3pd^3q}{(2\pi)^3(2p)^{1/2}(2q)^{1/2}}\Big{(}[a(\vec{p}),a^{\dagger}(\vec{q})]e^{-ip\cdot x+iq\cdot y}
+[a^{\dagger}(\vec{p}),a(\vec{q})]e^{ip\cdot x -iq\cdot y}\Big{)}
\nonumber\\
&&=\int \frac{d^3p}{(2\pi)^32p}\big{(}e^{-ip\cdot (x-y)}-e^{ip\cdot (x-y)}\big{)}
=-\frac{i}{2\pi}\frac{\delta(x^0-y^0-|\vec{x}-\vec{y}|)-\delta(x^0-y^0+|\vec{x}-\vec{y}|)}{2|\vec{x}-\vec{y}|}.
\label{Q10}
\end{eqnarray}
We note that this $i\Delta(x-y)$ is a c-number, and not a q-number, with (\ref{Q2}) following from (\ref{Q10}) since
\begin{eqnarray}
\frac{\partial}{\partial y_0}i\Delta(x-y)\big{|}_{x^0=y^0}=i\delta(x^1-y^1)\delta(x^2-y^2)\delta(x^3-y^3).
\label{Q11}
\end{eqnarray}
Since the unequal time $i\Delta(x-y)$ is defined at all $x^{\mu}$ and $y^{\mu}$, it is equally defined at equal light-front time $x^+=x^0+x^3=y^0+y^3=y^+$. It is the purpose of this paper to show that at equal light-front time (\ref{Q10}) precisely coincides with (\ref{Q6}). Since the form for the unequal instant-time commutator follows solely from the imposition of the equal-time commutator given in (\ref{Q2}) and the wave equation $\partial_{\mu}\partial^{\mu}\phi=0$ obeyed by the scalar field, the identification  of (\ref{Q10}) with (\ref{Q6}) would then entail that equal light-front time quantization is a consequence solely of equal instant-time quantization, with the light-front formulation not requiring any independent quantization of its own. Rather, it is just a consequence of instant-time quantization.  In this paper we will also obtain similar results for fermions and gauge bosons. Thus in all these cases light-front quantization is instant-time quantization. 

\section{Equivalence for Scalar Fields}
\label{S2}

To show the equivalence of instant-time quantization and light-front quantization in the scalar field case we first rewrite (\ref{Q10}) in the manifestly covariant form
\begin{eqnarray}
i\Delta(x-y)=-\textstyle{\frac{i}{2\pi}}\epsilon(x^0-y^0)\delta[(x^0-y^0)^2-(x^1-y^1)^2-(x^2-y^2)^2-(x^3-y^3)^2].
\label{Q12}
\end{eqnarray}
We set $x^0=\textstyle{\frac{1}{2}}(x^++x^-)$, $x^3=\textstyle{\frac{1}{2}}(x^+-x^-)$, $y^0=\textstyle{\frac{1}{2}}(y^++y^-)$, $y^3=\textstyle{\frac{1}{2}}(y^+-y^-)$ in (\ref{Q12}) and obtain
\begin{eqnarray}
i\Delta(x-y)=-\textstyle{\frac{i}{2\pi}}\epsilon[\textstyle{\frac{1}{2}}(x^++x^--y^+-y^-)]\delta[(x^+-y^+)(x^--y^-)-(x^1-y^1)^2-(x^2-y^2)^2].
\label{Q13}
\end{eqnarray}
Since $\epsilon(x/2)=\epsilon(x)$ for any $x$, at $x^+=y^+$ (\ref{Q13}) takes the form
\begin{eqnarray}
i\Delta(x-y)\big{|}_{x^+=y^+}=-\textstyle{\frac{i}{2\pi}}\epsilon(x^--y^-)\delta[(x^1-y^1)^2+(x^2-y^2)^2].
\label{Q14}
\end{eqnarray}
Then since $\delta(a^2+b^2)=\textstyle{\frac{\pi}{2}} \delta(a)\delta(b)$ for any $a$ and $b$, we can rewrite (\ref{Q14}) as \cite{Harindranath1996,Mannheim2019c}
\begin{equation}
i\Delta(x-y)\big{|}_{x^+=y^+}=[\phi(x^+,x^1,x^2,x^-), \phi(x^+,y^1,y^2,y^-)]
=-\textstyle{\frac{i}{4}}\epsilon(x^--y^-)\delta(x^1-y^1)\delta(x^2-y^2).
\label{Q15}
\end{equation}
We recognize (\ref{Q15}) as (\ref{Q6}), with the equal light-front commutation relation (\ref{Q6}) thus being derived starting from the unequal instant-time commutation relation (\ref{Q10}). Since the unequal instant-time commutation relation (\ref{Q10}) itself follows from the equal instant-time commutation relation (\ref{Q2}), we see that the equal light-front time commutation relation (\ref{Q6}) follows directly from the equal instant-time commutation relation (\ref{Q2}) and does not need to be independently postulated.

\section{Equivalence for Fermion Fields}
\label{S3}

In instant-time quantization the free fermionic Dirac action is of the form
\begin{eqnarray}
I_D=\int d^4x\bar{\psi}(i\gamma^{\mu}\partial_{\mu}-m)\psi. 
\label{Q16}
\end{eqnarray}
The canonical conjugate of $\psi$ is $i\psi^{\dagger}$, and the canonical anticommutation relations are of the form
\begin{eqnarray}
&&\Big{\{}\psi_{\alpha}(x^0,x^1,x^2,x^3),\psi_{\beta}^{\dagger}(x^0,y^1,y^2,y^3)\Big{\}}
=\delta_{\alpha\beta}\delta(x^1-y^1)\delta(x^2-y^2)\delta(x^3-y^3),
\nonumber\\
&&\Big{\{}\psi_{\alpha}(x^0,x^1,x^2,x^3),\psi_{\beta}(x^0,y^1,y^2,y^3)\Big{\}}=0.
\label{Q17}
\end{eqnarray}
When the fermion field obeys the Dirac equation $(i\gamma^{\mu}\partial_{\mu}-m)\psi=0$, the  on-shell Fock space expansion of the fermion field is of the form 
\begin{eqnarray}
\psi(\vec{x},x^0)=\sum_{s=\pm}\int \frac{d^3p}{(2\pi)^{3/2}}\left(\frac{m}{E_p}\right)^{1/2}[b(\vec{p},s)u(\vec{p},s)e^{-ip\cdot x}+d^{\dagger}(\vec{p})v(\vec{p},s)e^{+ip\cdot x}],
\label{Q18}
\end{eqnarray}
where $E_p=+[(p_1)^2+(p_2)^2+(p_3)^2]^{1/2}$, where $s$ denotes the spin projection, where the Dirac spinors $u(\vec{p},s)$ and $v(\vec{p},s)$ obey $(\slashed{p}-m)u(\vec{p},s)=0$, $(\slashed{p}+m)v(\vec{p},s)=0$, and where the non-trivial creation and annihilation operator anticommutation relations are of the form 
\begin{eqnarray}
\{b(\vec{p},s),b^{\dagger}(\vec{q},s^{\prime})\}=\delta_{s,s^{\prime}}\delta^3(\vec{p}-\vec{q}),\quad
\{d(\vec{p},s),d^{\dagger}(\vec{q},s^{\prime})\}=\delta_{s,s^{\prime}}\delta^3(\vec{p}-\vec{q}).
\label{Q19}
\end{eqnarray}
In terms of $i\Delta(x-y)$ as given in (\ref{Q10}), from (\ref{Q19}) we obtain the unequal instant-time anticommutator
\begin{eqnarray}
\big{\{}\psi_{\alpha}(x^0,x^1,x^2,x^3), \psi_{\beta}^{\dagger}(y^0,y^1,y^2,y^3)\big{\}}=\left[(i\gamma^{\mu}\partial_{\mu}+m)\gamma^0\right]_{\alpha\beta}i\Delta(x-y),
\label{Q20}
\end{eqnarray}
For the light-front case we set $\partial_0=\partial_++\partial_-$, $\partial_3=\partial_+-\partial_-$, and obtain
\begin{eqnarray}
\gamma^0\partial_0+\gamma^3\partial_3=(\gamma^0+\gamma^3)\partial_++(\gamma^0-\gamma^3)\partial_-
=\gamma^+\partial_++\gamma^-\partial_-,
\label{Q22}
\end{eqnarray}
with (\ref{Q22}) serving to define $\gamma^{\pm}=\gamma^0\pm \gamma^3$. In terms of $\gamma^+$ and $\gamma^-$ the Dirac action takes the form
\begin{eqnarray}
I_D=\textstyle{\frac{1}{2}}\int dx^+dx^1dx^2dx^-\psi^{\dagger}[i\gamma^0(\gamma^+\partial_++\gamma^-\partial_-+\gamma^1\partial_1+\gamma^2\partial_2)-\gamma^0m]\psi.
\label{Q23}
\end{eqnarray}
With this action the light-front time canonical conjugate of $\psi$ is $i\psi^{\dagger}\gamma^0\gamma^+$. 

In the construction of the light-front fermion sector we find a rather sharp distinction with the instant-time fermion sector. First, unlike $\gamma^0$ and $\gamma^3$, which obey $(\gamma^0)^2=1$, $(\gamma^3)^2=-1$, $\gamma^+$ and $\gamma^-$ obey $(\gamma^+)^2=0$, $(\gamma^-)^2=0$, to thus both be non-invertible divisors of zero. Secondly, the quantities 
\begin{eqnarray}
\Lambda^{+}=\textstyle{\frac{1}{2}}\gamma^0\gamma^{+}=\textstyle{\frac{1}{2}}(1+\gamma^0\gamma^3),\quad \Lambda^{-}=\textstyle{\frac{1}{2}}\gamma^0\gamma^{-}=\textstyle{\frac{1}{2}}(1-\gamma^0\gamma^3)
\label{Q24}
\end{eqnarray}
obey 
\begin{eqnarray}
\Lambda^{+}+\Lambda^{-}=I,\quad(\Lambda^{+})^2=\Lambda^{+}=[\Lambda^+]^{\dagger},\quad (\Lambda^{-})^2=\Lambda^{-}=[\Lambda^{-}]^{\dagger},\quad \Lambda^{+}\Lambda^{-}=0.
\label{Q25}
\end{eqnarray}
We recognize (\ref{Q25}) as a projector algebra, with $\Lambda^{+}$ and $\Lambda^{-}$ thus being non-invertible projection operators. Given the projector algebra we identify $\psi_{(+)}=\Lambda^+\psi$, $\psi_{(-)}=\Lambda^-\psi$ (respectively known as good and bad fermions in the light-front literature), and identify the conjugate of $\psi$ as $2i\psi_{(+)}^{\dagger}$, where $\psi^{\dagger}_{(+)}=[\psi^{\dagger}]_{(+)}=\psi^{\dagger}\Lambda^+=[\Lambda_+\psi]^{\dagger}=[\psi_{(+)}]^{\dagger}$. Since the conjugate is a good fermion, in the anticommutator of $\psi$ with its conjugate only $\psi_{(+)}$ will contribute since $\Lambda^+\Lambda^-=0$, with the equal light-front time canonical anticommutator being of the form (see e.g. \cite{Mannheim2019a} and references therein)
\begin{equation}
\big{\{}[\psi_{(+)}]_{\alpha}(x^+,x^1,x^2,x^-),[\psi_{(+)}^{\dagger}]_{\beta}(x^+,y^1,y^2,y^-)\big{\}}=\Lambda^+_{\alpha\beta}\delta(x^--y^-)\delta(x^1-y^1)\delta(x^2-y^2).
\label{Q26}
\end{equation}

In this construction the bad fermion $\psi_{(-)}$ has no canonical conjugate and is thus not a dynamical variable. To understand this in more detail we manipulate the Dirac equation $(i\gamma^+\partial_++i\gamma^-\partial_-+i\gamma^1\partial_1+i\gamma^2\partial_2-m)\psi=0$. We first multiply on the left  by $\gamma^0$ to obtain
\begin{eqnarray}
2i\partial_+\psi_{(+)}+2i\partial_-\psi_{(-)}+i\gamma^0(\gamma^1\partial_1+\gamma^2\partial_2)\psi-m\gamma^0\psi=0.
\label{Q27}
\end{eqnarray}
Next we multiply (\ref{Q27}) by $\Lambda^-$ and also multiply it by $\Lambda^+$ to obtain
\begin{equation}
2i\partial_-\psi_{(-)}=[-i\gamma^0(\gamma^1\partial_1+\gamma^2\partial_2)+m\gamma^0]\psi_{(+)},~~
2i\partial_+\psi_{(+)}=[-i\gamma^0(\gamma^1\partial_1+\gamma^2\partial_2)+m\gamma^0]\psi_{(-)}.~~
\label{Q28}
\end{equation}
Unlike the equation for  $\partial_+\psi_{(+)}$ with its $\partial_+$ derivative, the $\partial_-\psi_{(-)}$ equation contains no time derivatives. $\psi_{(-)}$ is thus a constrained variable, consistent with it having no conjugate.  Through the use of the inverse propagator $(\partial_-)^{-1}(x^-)=\epsilon(x^-)/2$ we can integrate the $\partial_-\psi_{(-)}$ equation in (\ref{Q28}) as  
\begin{equation}
\psi_{(-)}(x^+,x^1,x^2,x^-)=\textstyle{\frac{1}{4i}}\int dy^-\epsilon(x^--y^-)[-i\gamma^0(\gamma^1\partial_1+\gamma^2\partial_2)+m\gamma^0]\psi_{(+)}(x^+,x^1,x^2,y^-),
\label{Q29}
\end{equation}
and recognize $\psi_{(-)}$ as obeying a constraint condition that is  nonlocal. It is because $\psi_{(-)}$ obeys such a nonlocal constraint that it is known as a bad fermion. Since $\psi_{-}$ is a constrained variable it does not appear in any fundamental anticommutation relation, though one can use (\ref{Q26}) and (\ref{Q29}) to construct a non-local bad fermion equal light-front time anticommutator of the form 
\begin{eqnarray}
&&\Big{\{}\psi_{\mu}^{(-)}(x^+,x^1,x^2,x^-),[\psi_{(-)}^{\dagger}]_{\nu}(x^+,y^1,y^2,y^-)\Big{\}}=
\nonumber\\
&&\frac{\Lambda^-_{\mu\nu}}{16}\left[-\frac{\partial}{\partial x^1}\frac{\partial}{\partial x^1}-\frac{\partial}{\partial x^2}\frac{\partial}{\partial x^2}+m^2\right]
\int du^-\epsilon(x^--u^-)\epsilon(y^--u^-)\delta(x^1-y^1)\delta(x^2-y^2).~~~
\label{Q29a}
\end{eqnarray}

As with the analog scalar field case, starting from the unequal instant-time relation (\ref{Q20}) we now recover the light-front good and bad fermion equal $x^+$ anticommutators given in (\ref{Q26}) and (\ref{Q29a}) by transforming (\ref{Q20}) to light-front variables. For the good fermion first we multiply both sides of (\ref{Q20}) by $\Lambda^+$ on both the right and the left. Then on noting that
\begin{equation}
\Lambda^+\gamma^0\Lambda^+=0,~~ \Lambda^+\gamma^1\gamma^0\Lambda^+=0,~~ \Lambda^+\gamma^2\gamma^0\Lambda^+=0,~~ \Lambda^+\gamma^+\gamma^0\Lambda^+=0,~~ 
\Lambda^+\gamma^-\gamma^0\Lambda^+=2\Lambda^+,
\label{Q30}
\end{equation}
from the right-hand side of (\ref{Q20}) we obtain  \cite{Harindranath1996,Mannheim2019c}
\begin{eqnarray}
\Lambda^+_{\alpha\gamma}\left[(i\gamma^{\mu}\partial_{\mu}+m)\gamma^0\right]_{\gamma\delta}i\Delta(x-y)\Lambda^+_{\delta\beta}
=2i\Lambda^{+}_{\alpha\beta}\partial_{-}i\Delta(x-y).
\label{Q31}
\end{eqnarray}
We now substitute $x^0=\textstyle{\frac{1}{2}}(x^++x^-)$, $x^3=\textstyle{\frac{1}{2}}(x^+-x^-)$, $y^0=\textstyle{\frac{1}{2}}(y^++y^-)$, $y^3=\textstyle{\frac{1}{2}}(y^+-y^-)$, and using (\ref{Q13}) rewrite the right-hand side of (\ref{Q31}) as
\begin{eqnarray}
&&2i\Lambda^{+}_{\alpha\beta}\frac{\partial}{\partial x^-}i\Delta(x-y)
\nonumber\\
&&=\frac{1}{\pi}\Lambda^{+}_{\alpha\beta}\epsilon[\textstyle{\frac{1}{2}}(x^++x^--y^+-y^-)](x^+-y^+)\delta^{\prime}[(x^+-y^+)(x^--y^-)-(x^1-y^1)^2-(x^2-y^2)^2]
\nonumber\\
&&+\frac{1}{\pi}\Lambda^{+}_{\alpha\beta}\delta[\textstyle{\frac{1}{2}}(x^++x^--y^+-y^-)]\delta[(x^+-y^+)(x^--y^-)-(x^1-y^1)^2-(x^2-y^2)^2].
\label{Q32}
\end{eqnarray}
At $x^+=y^+$ (\ref{Q32}) takes the form
\begin{eqnarray}
&&2i\Lambda^{+}_{\alpha\beta}\frac{\partial}{\partial x^-}i\Delta(x-y)\big{|}_{x^+=y^+}=\Lambda^+_{\alpha\beta}\delta(x^--y^-)\delta(x^1-y^1)\delta(x^2-y^2).
\label{Q33}
\end{eqnarray}
Equating with the good fermion projection of the left-hand side of (\ref{Q20}) thus yields
\begin{eqnarray}
\Lambda^+_{\alpha\gamma}\big{\{}\psi_{\gamma}(x^+,x^1,x^2,x^-),\psi_{\delta}(x^+,y^1,y^2,y^-)\big{\}}\Lambda^+_{\delta\beta}&=&
\nonumber\\
\big{\{}[\psi_{(+)}(x^+,x^1,x^2,x^-)]_{\alpha},[\psi_{(+)}^{\dagger}]_{\beta}(x^+,y^1,y^2,y^-)\big{\}}&=&\Lambda^+_{\alpha\beta}\delta(x^--y^-)\delta(x^1-y^1)\delta(x^2-y^2).~~~~
\label{Q34}
\end{eqnarray}
We recognize (\ref{Q34}) as the light-front relation (\ref{Q26}). An analogous procedure leads to the equal light-front time bad fermion anticommutator given in (\ref{Q29a}) as well.

Thus in analog to the scalar field case, in the fermion field case we can construct both the good and bad equal light-front time anticommutators from the unequal instant-time anticommutators. And since the good and bad fermions combined span the full spinor space ($\Lambda^++\Lambda^-=I$) we recover all of the information contained in the light-front fermion sector. It thus follows that the two quantization procedures lead to the same physics, with matrix elements of products of fermion fields in the two cases consequently being equal, just as we showed in \cite{Mannheim2019a,Mannheim2019b}.

\section{Equivalence for Gauge Fields}
\label{S4}

For our purposes here it is convenient to take the instant-time gauge field action $I_G$ to be of the gauge fixing form 
\begin{eqnarray}
I_G=\int d^4x\left[ -\textstyle{\frac{1}{4}}F_{\mu\nu}F^{\mu\nu}-\textstyle{\frac{1}{2}}(\partial_{\mu}A^{\mu})^2\right]=\int d^4x\left[ -\textstyle{\frac{1}{2}}\partial_{\nu}A_{\mu}\partial^{\nu}A^{\mu}\right],
\label{Q35}
\end{eqnarray}
where $F_{\mu\nu}=\partial_{\nu}A_{\mu}-\partial_{\mu}A_{\nu}$ and $A_{\mu}$ is an Abelian gauge field. The presence of the $-\chi^2/2$ term where $\chi=\partial_{\mu}A^{\mu}$ causes $I_G$ to be neither gauge invariant  nor equal to the gauge invariant  Maxwell action $I_M=-\textstyle{\frac{1}{4}}\int d^4xF_{\mu\nu}F^{\mu\nu}$. 

Variation of the $I_G$ action with respect to $A_{\mu}$ yields an equation of motion of the form
\begin{eqnarray}
\partial_{\nu}\partial^{\nu}A_{\mu}=0.
\label{Q36}
\end{eqnarray}
The utility of using (\ref{Q35}) is that the various components of $A_{\mu}$ are decoupled from each other in the equation of motion.  Consequently, we can treat each component of $A_{\mu}$ as an independent degree of freedom, and apply the scalar field analysis given above to each one of them. In this formulation (\ref{Q36}) entails that $\partial_{\nu}\partial^{\nu}\chi=0$. If one imposes the subsidiary conditions $\chi(x^0=0)=0$, $\partial_0\chi(x^0=0)=0$ at the initial time $x^0=0$, then since $\partial_{\nu}\partial^{\nu}\chi=0$ is a second-order derivative equation it follows that the non-gauge-invariant $\chi$ is zero at all times.

Given (\ref{Q35}) one can define instant-time canonical conjugates of the form $\Pi^{\mu}=\delta I_{G}/\delta \partial_0A_{\mu}=-\partial^0A^{\mu}$. This then leads to equal instant-time commutation relations of the form 
\begin{eqnarray}
&&[A_{\nu},\Pi^{\mu}]=[A_{\nu}(x^0,x^1,x^2,x^3),-\partial^0A^{\mu}(x^0,y^1,y^2,y^3)]=-i\delta^{\mu}_{\nu}\delta(x^1-y^1)\delta(x^2-y^2)\delta(x^3-y^3),
\nonumber\\
&&[A_{\nu}(x^0,x^1,x^2,x^3),\partial_0A_{\mu}(x^0,y^1,y^2,y^3)]=ig_{\mu\nu}\delta(x^1-y^1)\delta(x^2-y^2)\delta(x^3-y^3),
\label{Q37}
\end{eqnarray}
and in analog to the scalar field case, to unequal instant-time commutation relations of the form 
\begin{eqnarray}
&&[A_{\nu}(x^0,x^1,x^2,x^3),A_{\mu}(y^0,y^1,y^2,y^3)]=ig_{\mu\nu}\Delta(x-y)
\nonumber\\
&&=-\textstyle{\frac{i}{2\pi}}g_{\mu\nu}\epsilon(x^0-y^0)\delta[(x^0-y^0)^2-(x^1-y^1)^2-(x^2-y^2)^2-(x^3-y^3)^2],
\label{Q38}
\end{eqnarray}
where $g_{\mu\nu}$ is the instant-time metric and $i\Delta(x-y)$ is the scalar field $i\Delta(x-y)$ as given in (\ref{Q12}) .

Given (\ref{Q35}) one can also define equal light-front time canonical conjugates of the form $\Pi^{\mu}=\delta I_{G}/\delta \partial_+A_{\mu}=-\partial^+A^{\mu}=-2\partial_-A^{\mu}$. This leads to equal light-front time commutation relations  of a form analogous to (\ref{Q5}) and (\ref{Q6}), viz. \cite{Mannheim2019a}
\begin{eqnarray}
&&[A_{\nu},\Pi^{\mu}]=[A_{\nu}(x^+,x^1,x^2,x^-), -2\partial_-A^{\mu}(x^+,y^1,y^2,y^-)]=-i\delta_{\mu}^{\nu}\delta(x^1-y^1)\delta(x^2-y^2)\delta(x^--y^-),
\nonumber\\
&&[A_{\nu}(x^+,x^1,x^2,x^-),\partial_-A_{\mu}(x^+,y^1,y^2,y^-)]=\textstyle{\frac{i}{2}}g_{\mu\nu}\delta(x^1-y^1)\delta(x^2-y^2)\delta(x^--y^-),
\nonumber\\
&&[A_{\nu}(x^+,x^1,x^2,x^-), A_{\mu}(x^+,y^1,y^2,y^-)]=-\textstyle{\frac{i}{4}}g_{\mu\nu}\epsilon(x^--y^-)\delta(x^1-y^1)\delta(x^2-y^2),
\label{Q39}
\end{eqnarray}
where $g_{\mu\nu}$ is now the light-front metric. Thus, in analog to the scalar field case, the last expression in (\ref{Q39})  follows directly from (\ref{Q38}), with the instant-time metric transforming into the light-front metric. Moreover, since the instant-time initial conditions have forced $\chi$ to be zero at all $x^0$, $\chi$ is thus zero at all $x^+$, with the subsidiary condition being maintained in the light-front case. Thus for gauge fields we again see that light-front quantization is instant-time quantization. The discussion for non-Abelian gauge fields is similar and is given in \cite{Mannheim2019a,Mannheim2019c}.

To conclude, we note that while we have only discussed the equivalence of instant-time and light-front time quantization for free fields, our results immediately apply to interacting theories. Specifically, since perturbative interactions cannot change a Hilbert space, once we show that the free instant-time and free light-front theories are in the same Hilbert space (as their commutators and anticommutators are related purely by kinematic coordinate transformations),  it follows that the interacting theories are in the same Hilbert space too. In fact in general we note that because of the general coordinate invariance of quantum theory, any two directions of quantization that are related by a general coordinate transformation must describe the same theory. Since the transformation $x^0\rightarrow x^0+x^3=x^+$ is one such transformation (a spacetime-dependent translation, and not a Lorentz transformation incidentally), it follows that light-front quantization is instant-time quantization.
The author would like to thank Dr. Luchang Jin for his helpful comments.


\begin{thebibliography}{99}
  
 \bibitem{Brodsky:1997de}\href{https://doi.org/10.1016/S0370-1573(97)00089-6}{S. J. Brodsky, H.-C. Pauli and  S.  S. Pinsky, Phys. Rept. \textbf{301}, 299 (1998).}
 
 \bibitem{Mannheim2019a} P. D. Mannheim, P. Lowdon and S. J. Brodsky, {\it Comparing light-front quantization with instant-time quantization}, SLAC-PUB-17390, in preparation.
  
 \bibitem{Mannheim2019b} \href{https://doi.org/10.1016/j.physletb.2019.134916}{P. D. Mannheim, P. Lowdon and S. J. Brodsky, Phys. Lett. B \textbf{797}, 134916 (2019).}

 \bibitem{Harindranath1996} \href{https://arxiv.org/abs/hep-ph/9612244}{A. Harindranath,	
\textit{An Introduction to light front dynamics for pedestrians}, arXiv:hep-ph/9612244v2.}
 
 
 \bibitem{Mannheim2019c}\href{https://arxiv.org/abs/1909.03548}{P. D. Mannheim, \textit{ 	
Light-front quantization is instant-time quantization}, arXiv1909.03548v1 [hep-ph].}

   

 
 \end{thebibliography}
\end{document}